**Limitations of Ordered Macroporous Battery Electrode Materials at High Charge and Discharge Rates**

Sally O'Hanlon[a], David McNulty[a], Ruiyuan Tian[b,c], Jonathan Coleman[b,c], and Colm O'Dwyer [a,c,d]*

[a] *School of Chemistry, and Tyndall National Institute, University College Cork, Cork, T12 YN60, Ireland*

[b] *School of Physics, Trinity College Dublin, Dublin 2, Ireland*

[c] *AMBER Centre, Trinity College Dublin, Dublin 2, Ireland*

[d] *Environmental Research Institute, University College Cork, Lee Road, Cork T23 XE10, Ireland*

* Email: c.odwyer@ucc.ie  (Colm O'Dwyer) Tel: + 353 (0)21 4902732.

**Abstract**

Adding porosity to battery electrodes is believed to be universally useful for adding space to accommodate volumetric expansion, electrolyte access to all active materials, helping to mitigate poor C-rate performance for thicker electrodes and for allowing infilling with other materials. When porosity is used together with reduced material dimensions, the motivation for investigating better battery performance cited nanoscale dimensions to reduce solid state diffusion in the active material, and overall improve the rate performance. Ordered porous electrode, such as inverse opals that have macroporosity, have been a model system: binder and conductive additive free, interconnected electrically, defined and controllable porosity and pore size consistent with thickness, good electrolyte wettability and surprisingly good electrode performance in half cells and some Li- battery cells at normal rates. We show that the intrinsic electronic conductivity is important, and at fast rates the intrinsic conductivity ultimately supresses any charge storage in electrode materials. Using a model system of inverse opal $V_2O_5$ in a flooded Li battery three-electrode cell, whose Li electrochemistry is very well understood, we show that beyond 10 C, electrodes can store almost no charge, but completely recover once reduced to < 1C. We show how the IO material is modified under lithiation using X-ray diffraction, Raman scattering and electron microscopy, and that little or no reaction occurs to the material at higher rates. We also use chronoamperometry to examine rate behaviour and link the limitations in high rate performance, and complete capacity suppression, to the intrinsic out-of-plane conductivity of the IO network. The data show that even idealised electrodes with nanoscale dimensions, functional porosity and full material interconnectivity, are fundamentally limited for high rate performance when they are less conductive even when fully soaked with electrolyte. While adding so-called functional size reduction, porosity etc. can be useful for some materials, these potential benefits are clearly not universally useful for high rate electrodes in Li-ion batteries.

## Introduction

The need for enhanced battery materials in portable electronic devices(1, 2) and electric vehicles(3, 4) is essential, and batteries that can deliver higher capacity and cycle stability at faster rates will be important for enabling the electric future of transportation(5, 6). The way in which battery materials are manufactured and modified has emphasized the need for control of size, composition and structural integrity during use.(7-10) Electrode materials and structural design is one route to enabling batteries to deliver power or energy quickly(11, 12), and store full capacity when charged quickly(13). However, above a certain threshold, the maximum achievable capacity for a particular applied current, begins to fall off rapidly(14), particularly for thicker electrodes. This effect is quite dramatic when comparing high rate behaviour of thick electrodes(15, 16) compared to low rate behaviour, where areal capacity is severely suppressed at higher rates(17).

The material-based approach to deal with rate limitation and to improve energy storage include synthesis methods that allow control over electrical conductivity(18-20), porosity and material dimension (important for solid state cation diffusion)(21, 22), and optimizing the choice of electrode slurry additive where relevant(23-25). Optimization of separator type(26) and electrolyte ionic conductivity are also important since lithium transference number and dilution of the active material that are controlled and affected by lithium salt diffusion to the active material in composite electrodes.

The structure of the active electrode material, and indeed the overall composite in slurry formulations, has been a research focus in newer energy storage materials(27), especially for higher capacity and higher voltage cathodes. This is partly because porosity, material interconnection to improve interparticle electrical conductivity(28), the use of small sizes for cation insertion with minimal solid state diffusion limitations, and the inclusion of conductive additives etc. have all tackled the limitations of good higher rate performance(29). Inverse opal networks of many materials have several benefits in principle, many of which have been proven or detailed elsewhere(30-32) at least as a model system that used interconnected material with defined material size and geometric porosity. This architecture is becoming increasingly more popular for several electrochemical technologies owing to the shorter diffusion distance, easier infiltration of electrolyte, along with a larger surface area of a continuous network of electrode material(33-35) in three-dimensions. Aside from the obvious reduction of volumetric energy density caused by porosity, fashioning thicker electrodes and eliminating binders and conductive additives can offset some of this mass-related energy density reduction. These structures do simplify interpretation of what limits high rate behaviour compared to some more disordered composite electrodes (8, 36-41). One reason is that thicker IOs with defined porosity are all filled with electrolyte, thereby putting an upper limit on Li-ion diffusion limitation irrespective of the electrode thickness. What does change is the out-of-plane conductivity. Creating thin films of active material on 3D porous current collectors helps in this regard, but the good high rate ability stems from fast ion diffusion into thin films in a porous structure with a shortened diffusion distance also in the electrolyte(42-44). It is becoming clear that engineering porous interconnected materials allows faster Coulombically efficient rates, but it is not fully clear whether electronic conductivity or electrolyte diffusion depletion dominate over solid state diffusion limitation and kinetics of reactivity in general. For instance, some porous materials may be more electrically resistive, others react with lithium using other mechanisms (intercalation, alloying etc.) and intra-electrode diffusion can depend on the material porosity, type and crystal structure(39). For electrolytes, maximum infiltration is possible, so that salt diffusion versus electronic conductivity and solid-state diffusion contributions can be addressed. Park et al. recently developed a quantitative model to relate the areal capacity of electrodes with known thickness, to the rate capability and this provided a prediction for optimizing the trade between high areal capacity and the maximum rate performance(17). Electrical conductivity and porosity are critical for fast rate electrodes to maintain high capacity, more so than solid state diffusion limitations and reaction kinetics in some cases. Recently, several papers have focused on chronoamperometry and related methods to obtain rate response information from composite electrodes in a much shorter time frame compared to galvanostatic methods. Tian et al.(14) and Huebner et al. have proposed very useful quantitative methods(45, 46) to quickly obtain rate performance from low (~0.01 C) to high (>$10^6$ C) rates from current transients polarised at the electrodes lower cut-off potential. A consensus is being found where electrolyte diffusion limitations and electrode conductivity are critical for good high rate performance, especially in thicker electrodes(47). This approach has not yet been used to assess high rate response in binder and carbon-additive-free ordered porous electrodes.

It has been difficult to define a general understanding of the influence of material type and porosity under identical conditions, especially since porosity can be suppressed within calendared slurries compared to other composites. Some electrodes do not include additives or post-deposition modification. We developed 3D porous inverse opal networks of the cathode material $V_2O_5$ by electrodeposition in order to investigate the rate performance of this structure and material, without any additives or cell assembly methods to gauge its intrinsic response. Electrodeposition of active materials is especially useful for infilling of the opal template.

By ensuring initial deposition onto the current collector via nucleation with subsequent deposition upwards throughout the porous template, a high quality IO network of material in a crystalline, electrically interconnected structure is created(34). Vanadium pentoxide has been a popular candidate for studying Li intercalation into layered materials, offering a reasonably high volumetric and areal capacity and well-defined, step-like voltage profile in common organic electrolytes. $V_2O_5$ benefits in numerous ways such as its mixed valence allowing for unit cell changes during lithiation(7, 48). Being low cost and a relatively abundant source, $V_2O_5$ also has a layered van der Walls structure that accommodates Li-ion insertion and removal(49). The Li chemistry with $V_2O_5$ is well known(50-52), both in nanomaterials studies, thin film investigations and in slurry composites(53-56), and more recently as ordered macroporous or inverse opal structures(9). It is seeing a resurgence in Li-ion battery science(57) because of these traits, and becoming a candidate for Na-ion(58), Zn-ion systems(58) and very recently, as a porous electrode formulation for Al-ion batteries(59). The prior knowledge of lithiated vanadate electrochemistry is useful when trying to elucidate the influence of the ordered porous structure on rate performance.

With this 3D $V_2O_5$ IO material, we analysed the electrochemical response as a function of rate in detail. We first detail the growth of the IO electrodes by electrodeposition and how overfilling of porous IO materials occurs in tandem with the thickening of the IO cathode. The capacity, lithiation mole fraction, phase changes and energy density are compared at slow and fast rates. We find that thin, open IO structures behave well at lower rates, whereas at higher rates (>10 C) we show that capacity can be almost entirely suppressed in the regime where electrical conductivity and porosity are more important. At faster rates, thicker IO electrodes had lower overall capacity values (even when infilled in a flooded cell) but both IO electrodes exhibited near zero capacity at 30 C, which fully recovered when cycling continued at low rate. Correspondingly, no changes to the structure from lithiation occurred at these rates, implying almost no reaction. While lithiated vanadates are not very electrically conductive, we provide data and an interpretation using a recently developed model for quantifying rate behaviour in battery electrodes, to show that the intrinsic nature of IO oxide cathodes limits high rate behaviour and that introducing porosity per se must be tuned to ensure that capacity and rate are optimized for a given material.

**Experimental section**

*Electrodeposited $V_2O_5$ inverse opal structures*

Colloidal crystal templates were formed on a conductive substrate of FTO coated glass of ~1 cm × 1 cm geometric area by drop-casting of 0.5 μm diameter polystyrene (PS) spheres. The PS spheres from Polysciences Europe GmbH were functionalized with sulfate groups to aid in self-assembly. Vanadium pentoxide was infilled by electrodeposition at room temperature using a VersaSTAT3 Potentiostat. A potential of 2 V was applied for 300 s (open-IO surface) and 900 s (closed-IO surface) versus a saturated calomel (SCE) reference electrode in a three-electrode cell with a platinum mesh as counter electrode. The PS template-coated substrate was used as the working electrode. The electrolyte was made by adding 2.53 g of $VOSO_4 \cdot \chi H_2O$, used as purchased from Sigma Aldrich, to a 1:1 (v/v) mixture of 20 ml of deionized water and 20 ml of ethanol to form a 0.25 mol dm$^{-3}$ $VOSO_4 \cdot \chi H_2O$ solution(60). After deposition, samples were heated to 300°C for 24 h to remove the spheres, resulting in the formation of an inverse opal network of crystalline $V_2O_5$.

*Electrochemical characterization*

The electrochemical properties of the 3D $V_2O_5$ IO structures were investigated using a three-electrode cell using Biologic SP-150 and VMP3 systems. The cells were assembled inside a glovebox under an argon atmosphere. The electrolyte consisted of a 1 mol dm$^{-3}$ solution of $LiPF_6$ salt in a 1:1 (v/v) mixture of ethylene carbonate (EC) in dimethyl carbonate (DMC), with 3 wt% vinylene carbonate (VC) as an electrolyte additive. No additional conductive additives or binders were added to the $V_2O_5$ IO working electrodes, allowing direct electrochemical examination of each structure without contribution from conductive additives, binders and non-uniform mixtures. The cell was setup as a flooded cell, allowing full infiltration of the IO electrode materials with electrolyte. To avoid inconsistent compression difference between electrodes and to eliminate any contributions from compressed or pulverised IO electrode materials, the flooded cell was setup to eliminate downward pressure on the electrode typically found in an assembled coin cell.  The counter electrode and a separate reference electrode were pure lithium metal. Galvanostatic and rate capability testing was performed using a range of different C-rates between 0.5 C and 30 C, where in this work 1 C = 147 mA g$^{-1}$ (at $\chi \sim 1$ in $Li_\chi V_2O_5$), in a potential window of 4.0−1.2 V. Chronoamperometry was performed using fresh IO electrodes in the same 3-electrode cell used for galvanostatic charge-discharge data described above. The potential was stepped from open circuit to 1.2 V (lower cut-off for discharge curves) without any pre-lithiation. The method by which the $I(t)$ transient is converted to a relative specific capacity vs rate (either R-rate or C-rate) are outlined in several papers.(14, 17, 46, 61-63)

*Materials characterization and analysis*

Structural and morphological characterization of the electrochemically tested $V_2O_5$ IO structures was performed using a Zeiss Supra 40 SEM at accelerating voltages in the range 5–10 kV. Crystallographic information was investigated using Raman scattering spectroscopy and X-ray diffraction. Raman scattering was acquired using a QE65 PRO Ocean Optics spectrometer with a 50 μm slit width. Excitation was provided by a Laser Quantum GEM DPSS single transverse mode CW green laser emitting at $\lambda$ = 532 nm. The spectral resolution of the spectrometer ranges from 17.5–10.5 cm$^{-1}$ between 300–4000 cm$^{-1}$. The laser source was focused onto the sample surface through an objective of 4×, 10×, 20× or 40× magnifications with numerical apertures (N.A.) of 0.10, 0.25, 0.40 and 0.60, respectively. Spectra were collected under a variety of different laser powers from 10 mW to 100 mW. The laser power densities (LPDs) calculated from these settings range between 15.41 W cm$^{-1}$ for 4× magnification with 10% laser power to 924.3 W cm$^{-1}$ for 40× magnification with 100% laser power, and care was taken to avoid localised heating that affected phonon modes in the spectra. X-ray diffraction was performed on a Philips Xpert PW3719 diffractometer using Cu Kα radiation.

**Results and Discussion**

*Electrodeposition and structure of $V_2O_5$ inverse opal electrodes*

Electrodeposition provides a simple and straightforward method for growing inverse opal (IO) oxides with good control over the resulting mass under potentiostatic conditions(64). Summarised in Figure 1 is the general process for forming crystalline $V_2O_5$ in IO format (Figure 1(a,b)) directly onto current collectors from the colloidal photonic crystal PS template. The ordered and seemingly smooth walled structured arises from the layered $V_2O_5$ morphology. TEM images confirm the orientation of the layers within these structures. In octahedral voids(65, 66), the layering spans across the gap, while the walls maintain the layered structure. One of the assumptions for macroporous or 'functional' porous materials in Li-ion battery electrode research, is the benefit of shorter Li-ion diffusion distance within the solid (thin walls) and a fixed localized volume of electrolyte with each void(43, 44, 67, 68). The materials are also continuously interconnected, electrically conductive and, in principle, also provides a buffer for volume expansion and delamination. In this work, the battery cells are flooded, mimicking standard three-electrode cells using Li as a reference electrode to allow careful measurement of the lithium battery cell voltage. The electrolyte infills all pores within the IO structure. As there are no binders, issues associated with binder overlayers formed on composite cast films during drying are eliminated entirely. Ex situ examination of porous IO liquid soaking (dropwise addition of solvent to the top open surface) confirmed that solvents easily penetrate and soak into the IO porous material, without causing changes to the material structure. Microporosity-induced tortuosity(69), for fully connected and dead-end pores within a complex slurry composite electrode, are replaced by ordered macroporosity where cationic tortuosity is negligible.

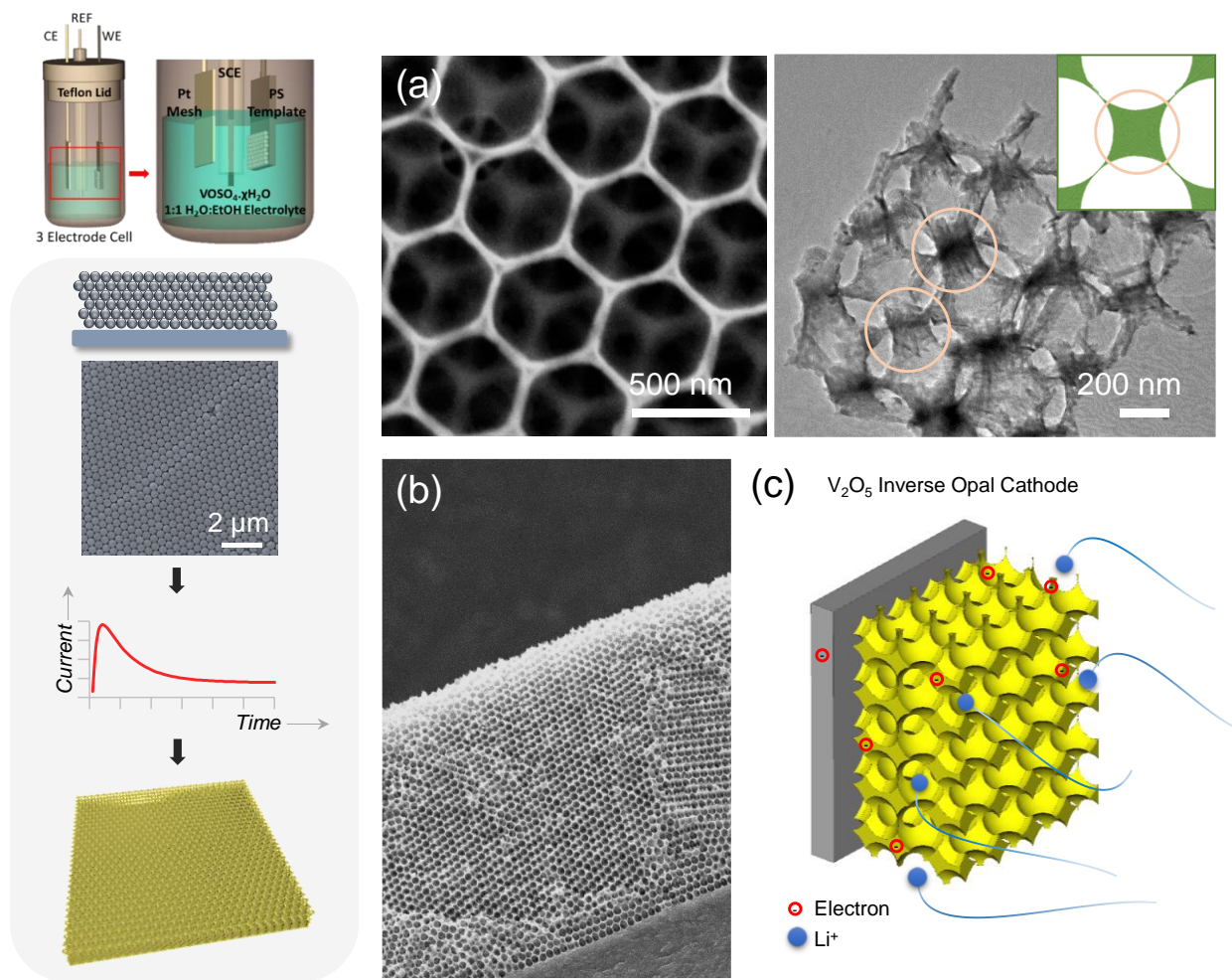

**Figure 1.** Schematic of the electrodeposition setup to form crystalline inverse opals of $V_2O_5$ directly onto current collectors template by an opal colloidal photonic crystal comprising 500 nm diameter PS spheres. (a) SEM and TEM images of a typical $V_2O_5$ IO structure. $V_2O_5$ as a van der Waals (vdW) layered material exhibits characteristic layering within the wall of the IO. The vdW layers are host gallery spacings for Li$^+$ during intercalation reactions. (b) SEM of a thick IO structure electrode with (c) a representation of the Li intercalation process as a function of out-of-plane IO electronic conductivity and void-filled electrolyte Li-ion diffusivity.

Electrodeposition creates a crystalline vdW layered orthorhombic phase directly on the current collector, and the open ordered porous network has been shown to improve reversible rate-dependent lithiation as a Li-battery cathode than a corresponding V$_2$O$_5$ thin films(9). These rates, like many other studies, are not excessively high and likely did not test the limitations of material conductivity and ionic diffusion. With a measured electronic conductivity of ~10$^{-4}$ S cm$^{-1}$, there has been little examination of the effect of IO electrode thickness and overfilling (where the IO has an outer covering film of the same material) on the nature of the lithiation response(39) for structured interconnected porous materials at high C-rates. SEM data in Figure 2(a) shows the structure of very high quality interconnected V$_2$O$_5$ IO structures, and for short times (300 s) are consistent with previous reports with IO pore diameter ranging from 465-480 nm.(9, 34) These structures were electrodeposited into opal photonic crystal templates using 500 nm PS spheres and the electrodeposition is a diffusion-limited process(34, 60) leading to a constant integrated charge or growth rate per unit time after initial surface nucleation on the substrate. As IO materials are expected to provide enhanced electronic and ionic mobility enabling faster charge and discharge rates to full capacity, we decided to investigate the effect of thicker IO films and overfilling of the porous, electrolyte-accessible top surface on the response to faster charge rates in lithium batteries.

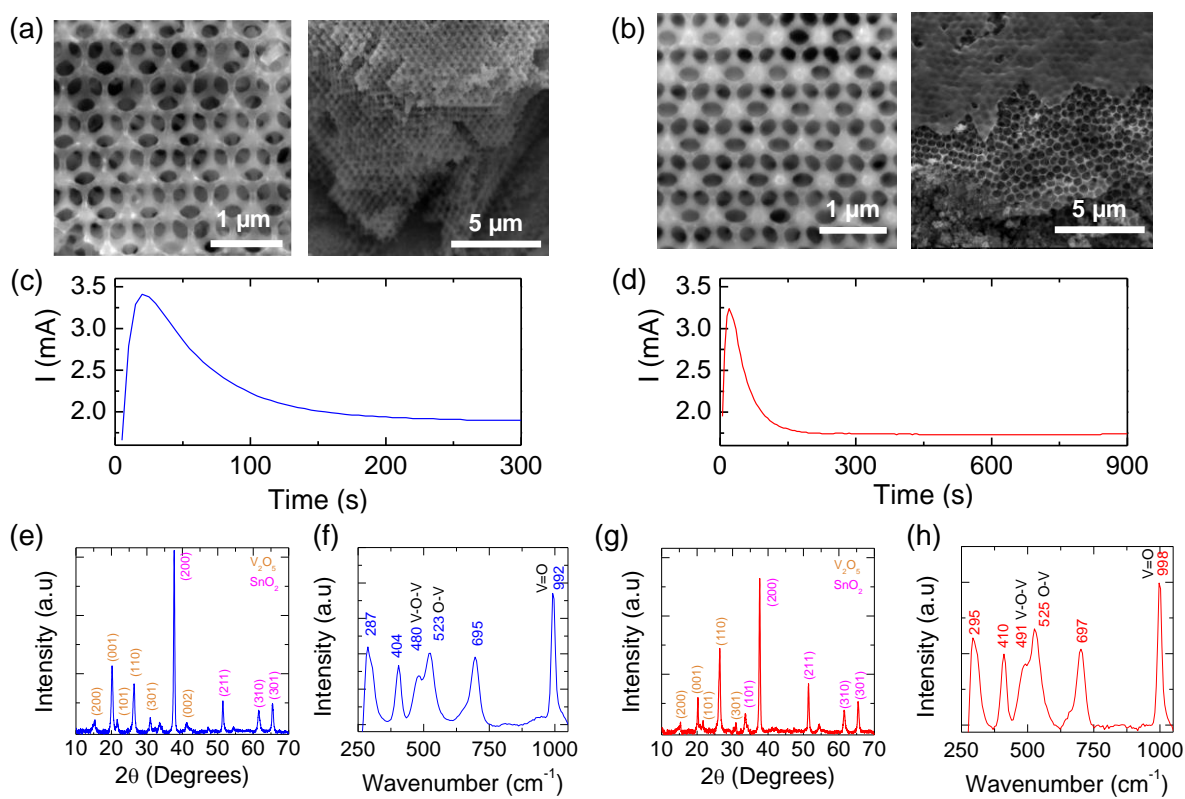

**Figure 2.** (a) SEM images of V$_2$O$_5$ IO grown by electrodeposition at a potential of 2 V (vs SCE) for 300 s and (b) for 900 s. Tilted images confirm IO formation throughout the original opal template and the overlayer characteristic of longer growth times. We refer to these as open IO (o-IO) and closed IO (c-IO). (c) Corresponding potentiostatic $I$(t) chronoamperograms for growth of V$_2$O$_5$ IO for 300s and (d) 900s. (e,g) XRD patterns of V$_2$O$_5$ indexed to PDF 41-1426 and (f,h) Raman scattering spectra for IO deposits following growth for 300 s and 900 s respectively. Reflections from the underlying FTO substrate (PDF 46-1088) are marked.

The open structure formed after 300 s electrodeposition is referred to as the open IO (o-IO). These structures are formed by partial infilling of the IO template, followed by standard template removal to provide a thinner IO of ~5 – 7 µm. After electrodeposition for 900 s, SEM images (Figure 2(b)) confirm an ordered porous interconnected 3D structure. The top surface is partly overfilled and electrodeposition for 900 s results in a dense film of $V_2O_5$ partially covering the IO. A period of 900 s was chosen to ensure that the fixed opal template thickness of (~10 – 15 layers of 500 nm sphere ≈ 9 – 12 µm) was completely filled. The overfilled regions likely arise from regions of the opal template that were thinner, which is a common occurrence for dip-coated and drop-case opal templates. Even a difference of 1-2 layers of spheres from a template of 10-15 layers of 500 nm spheres will cause overfilling or pooling from electrodeposition in those local top surface regions. This is referred to as a closed-IO (c-IO) electrode in this work. This type of overfilling is more common in IO formation than is typically reported in the literature and the influence of irregularities in coverage of materials designed to add functionality using porosity, is rarely assessed.

The current (*I*) - time (*t*) chronoamperograms in Figure 2 (b) for electrodeposition over periods of 300 s and 900 s, each show a similar growth pattern where the initial sharp rise from instantaneous nucleation, which then falls and levels to a steady (linear) growth at just under 2 mA in each case. The diffusion coefficient determined(70-72) using the Cottrell relation ($I - t^{1/2}$) values at 2.0 V is $6.1 \times 10^{-7}$ cm$^2$ s$^{-1}$ and as previously determined, are not markedly affected by the tortuosity of the ionic diffusion through the opal template. For instance, we observed an order of magnitude increase in ionic diffusion rate to the surface of ~$6.5 \times 10^{-6}$ cm$^2$ s$^{-1}$ with no template in place. From SEM data, full infiltration of the voids in the opal template is achieved by electrodeposition. We consistently observed at this potential, a non-zero instantaneous current. The non-zero instantaneous current for such a nucleation mechanism is indicative of the free open area between the first layer of PS spheres in the template and the current collector. Based on geometrical considerations(73-77) of the fcc packing factor of the opal template, the total surface area for nucleation under the first layer of spheres is $A_{ED} = \frac{V_{ED}}{L_{x,y}}$, where $L_{x,y} = n\Lambda$ for *n* spheres and the distance parameter is $\Lambda = \frac{L_{x,y}}{n} = \frac{\phi}{n}$ where $\phi$ is the PS sphere diameter. The effective open area for electrodeposition in this case is consistently $\frac{\pi}{6\sqrt{3}}$ ~ 0.302 cm$^2$ of a 1 cm$^2$ substrate. Using the density of $V_2O_5$ of 3.36 g cm$^{-3}$ and a porosity of 74% for an ideal IO, theoretical mass values are 0.15 and 0.3 mg for the o-IO and c-IO, respectively. These values are close to measured values in the range 0.22 – 1.1 mg for o-IO and c-IO used for all electrodes in this work, from which corresponding specific currents (C-rate) were determined. The larger masses likely accommodate the dense overfilling of the c-IO thicker samples.

X-ray diffraction date (Figure 2(e,g) and Raman scattering spectra (Figure 2(f,h)) confirm orthorhombic $V_2O_5$ (Space Group P*mmn*) for each $V_2O_5$ IO structure. For electrodeposition over 900 s where $V_2O_5$ is grown beyond the thickness of the opal template leading to an overfilling by a dense film coating (c-IO), the (110) reflection is more intense. This specific diffraction intensity change is due to a thin film layer on top of the IO structure. In the IO, the (110) planes are confined to thin dimensions in the octahedral voids mainly. In the overfilled film, the volume fraction of (110) planes is not spatially limited over the effective surface area $A_{ED}$ of the substrate. Filming of crystalline $V_2O_5$ predominantly grows with (110) planes parallel to the underlying substrate or layer.

*Rate behavior of $V_2O_5$ inverse opal Li battery cathodes*

Both types of $V_2O_5$ IO cathodes were subjected to galvanostatic testing at various C-rates to compare electrochemical response, relative performance and how the IO material behaves. Specifically, we sought to identify any limits to faster discharge and charge rates for IO structures and determine if solid state cation diffusion rates, electrolyte cation diffusion from the electrolyte to the surface, and the overall IO conductivity are intrinsic to performance limits. A variety of C-rates were tested from slower rates of 0.5 C to faster rates of 30 C. The 1st discharge profiles for both o-IO and c-IO structures are shown in Figure 3. The distinct stepped voltage profile this is well known for the first discharge of a $V_2O_5$ cathode is observed in both cases, showing the phase changes associated with very well-defined lithiation to higher Li mole fractions(7) in $Li_xV_2O_5$. The o-IO $V_2O_5$ Li battery cell delivers a specific capacity of ~302 mAh g$^{-1}$ at 0.5 C.

When the IO electrodes in the flooded Li battery cells are subjected to faster C-rates of 5 C, 10 C and 30 C, the IO structure maintains the characteristic voltage steps (phase changes) in the lithiation process during the 1st discharge at each rate. Specific capacity values reduce with higher C-rates as expected to values of ~243, 81 and 60 mAh g$^{-1}$, respectively for the o-IO structure. An order of magnitude increase in lithiation rate from 0.5 C to 5 C reduced the capacity recorded in the 1st discharge of a freshly made o-IO cell by just ~18%. As the structure was discharged at faster rates of 10 C and 30 C, the voltage profile is significantly different where we see a significant reduction in specific capacity by 75% from a doubling of rate from 5 C to 10 C. As will be shown in more detail later, the voltage steps remain for o-IO structures at these high rates and occur at similar potentials at all rates examined. This confirms that some lithiation processes occur at the faster rates.

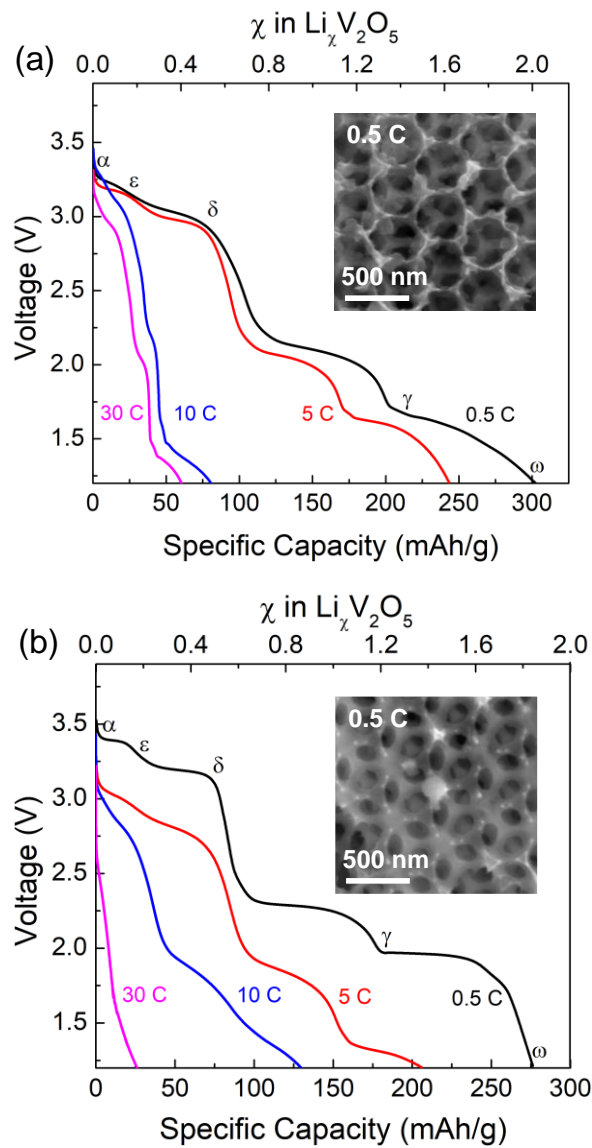

**Figure 3.** Galvanostatic 1st discharge curves for $V_2O_5$ IO electrodes in lithium batteries at various C-rates in the range 0.5 – 30 C. The cathode materials were formed by electrodeposition at 2.0 V (vs Li+/Li) for (a) 300 s and (b) 900 s. The definitive phase changes to the orthorhombic $V_2O_5$ crystal structure during lithiation are marked α - ω. The insets show electron microscopy images of the internal IO network after discharge to 1.2 V (vs Li+/Li) at a rate of 0.5 C. The image shown in (b) is of the c-IO electrode that was not covered with the dense $V_2O_5$ overlayer.

For the thicker c-IOs, the voltage plateaus of the lithiation process were clearly observed for the 0.5 C and 5 C discharge rates, however as the C-rate increased to 10 C and 30 C, these specific phase changes became less pronounced. The specific capacities were 276 mAh g$^{-1}$ (0.5 C), 206 mAh g$^{-1}$ (5 C), 130 mAh g$^{-1}$ (10 C), and 26 (30 C) mAh g$^{-1}$. A characteristic difference is seen in the voltages of the lithiated vanadate phases, which required more energy to transition to higher mole fractions, indicated by the lower voltages. This, we believe, is due to lithiation of the dense thin film material covering the c-IO in addition to the IO material. The fully lithiated a thin film partially covering the IO requires a voltage penalty to retain the specific current associated with each phase change (high lithium mole fraction) compared to the thinner IO walls. Inset SEM images in Figure 3 confirm that the structures are maintained after the slower (0.5 C) rate. The primary difference is that the internal wall structure of the o-IO was clearly modified from lithiation, compared to the c-IO whose internal structure was less affected by volumetric expansion and surface roughening. Lithiation of the dense film on the IO outer surface contributed to the overall specific capacity for the c-IO material.

The differential charge ($\frac{dQ}{dV}$) curves for each of the $V_2O_5$ IO structures is presented in Figure 4(a) using the 1$^{st}$ discharge data, and the associated potentials for all phases are show in Figure 4(b) as a function of C-rate. The shift of the differential peaks to lower voltages for all lithiated vanadate phase changes is consistent with diffusion-limited systems. At higher C-rates, more energy is needed to induce lithiation to match the specific current (reaction rate) during lithiation causing a change in voltage associated with known phase transitions. From Figure 4(a) and Figures 4(b,d), phase changes in the o-IO are well pronounced – as phase changes are observed even up to 30 C where the initial discharge capacity is quite low. We do find that the irreversible ω-Li$_x$V$_2$O$_5$ phase dominates the contribution to the low overall specific capacity.

For c-IOs (Figure 4(c) and Figures 4(b,d)), phase changes of the δ, γ, and ω-Li$_x$V$_2$O$_5$ are shifted to lower voltage and the lithiation mole fraction associated to each phase occurs over a broader potential range. The ω phase for the o-IO structure starts at 1.698 V at the 0.5 C rate and decreases to 1.371 V at 30 C, whereas the ω phase in the closed-top structure shifts from 1.976 V to 1.251 V. Likewise, the γ phase for the o-IO starts at 2.246 V at the 0.5 C rate and decreases to 2.056 V at 30 C, whereas the γ phase in the c-IO reduces from 2.295 V to 1.567 V. And finally, the δ phase for the o-IO starts at 3.152 V at the 0.5 C rate and decreases to 2.968 V at 30 C, whereas the δ phase in the closed structure drops from 3.212 V to 1.813 V. In Figure 4(d), we compare the corresponding lithiated phase (lithium mole fraction) with C-rate for both o-IO and c-IOs. The δ-Li$_x$V$_2$O$_5$ phase underwent the largest reduction in voltage at higher rates when the IO was thicker and partly overfilled with a dense thin film.

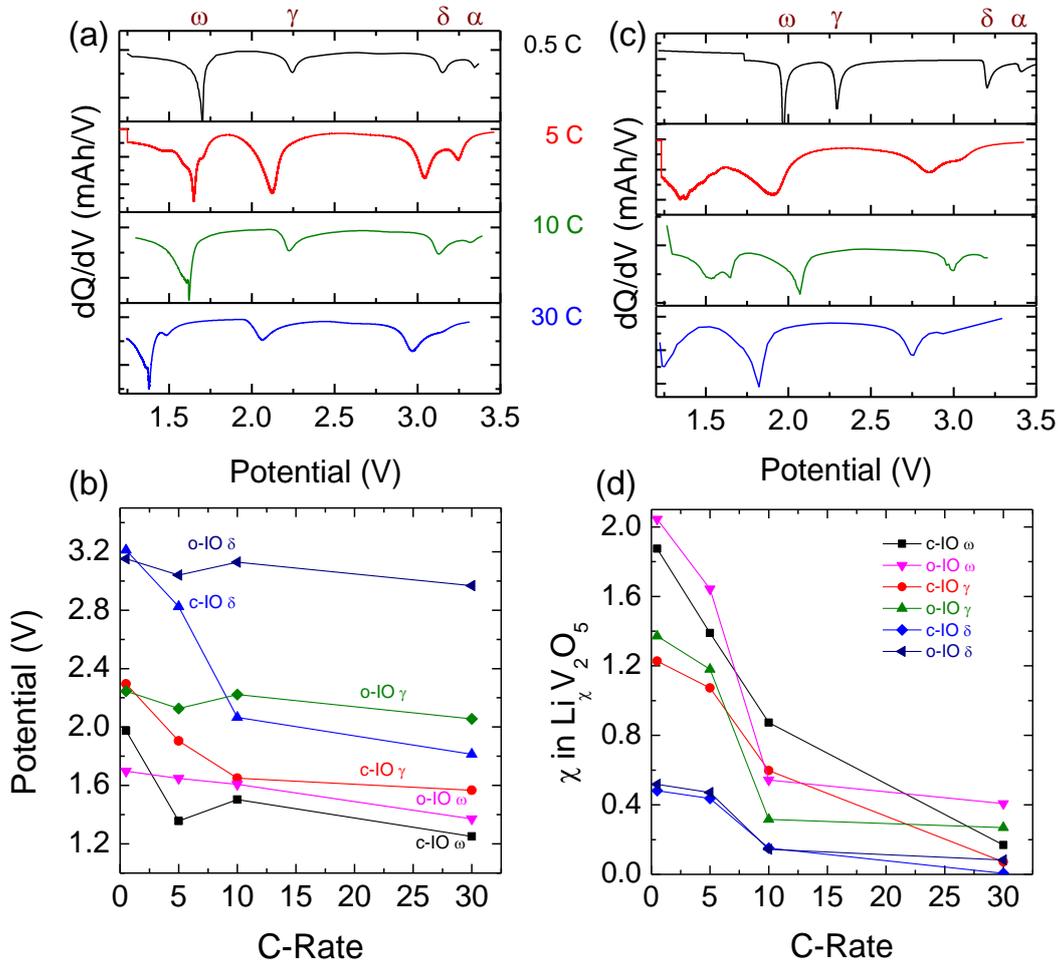

**Figure 4.** Differential charge curve for (a) o-IO formed by ED for 300 s and (c) for the c-IO grown over 900 s, from the first galvanostatic discharge of each C-rate. The corresponding $Li_xV_2O_5$ phases are indicated. (b) Potential (V) vs C-rate of each $Li_xV_2O_5$ phase for $V_2O_5$ o-IO and c-IO structures. (d) Lithium mole fraction ($\chi$) vs C-rate for both open (o-IO) and closed-top (c-IO) $V_2O_5$ IO structures from the first discharge.

Compared to the o-IO, the overlayer and thickness had comparatively negligible effect on the degree of lithiation for this lithiated vanadate phase. It is the o-IO structure that is most affected by higher rates, and we can track the capacity reduction at high rates to the lower mole fractions associated with $\gamma$ and $\omega$-$Li_\chi V_2O_5$ at rates close to 10 C. Both electrodes are nominally filled with electrolyte in our flooded cells, and so issues with local electrolyte depletion or wetting issues are ruled out.

Separate lithium battery flooded cells were also tested using the o-IO and c-IO cathodes. From the discharge curves, the energy density at each voltage step or lithium mole fraction-related phase change was determined and shown in Figure 5. In each case, the energy densities reduced with higher C-rate as would be expected. The o-IO structures showed a similar reduction in the relative energy density related to each phase as a function of C-rate, but the c-IO was comparatively less consistent.

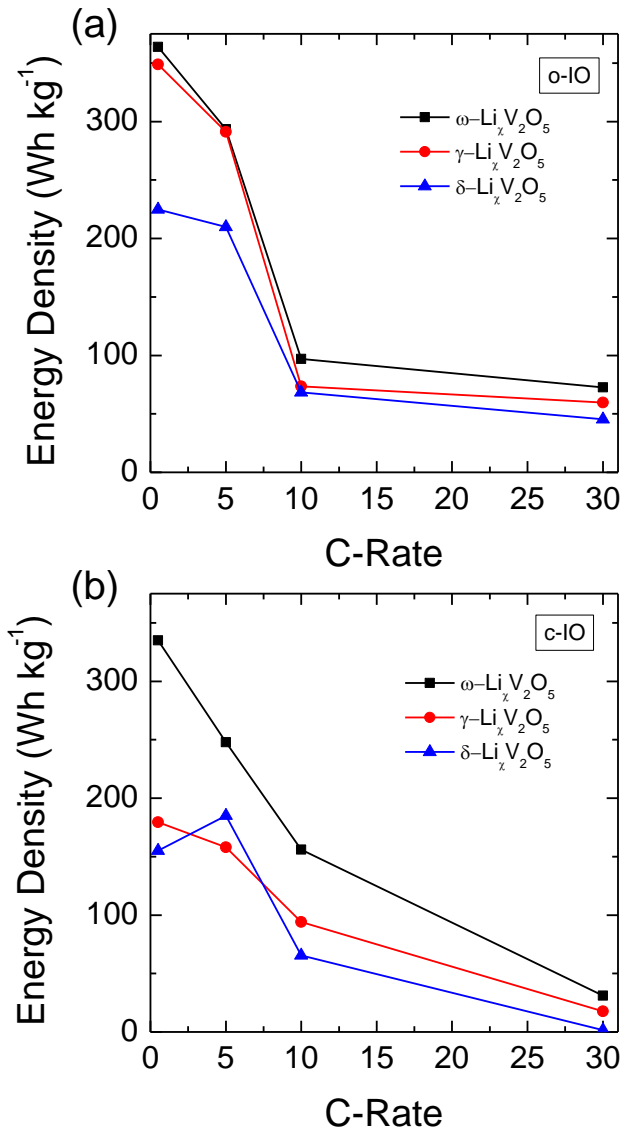

**Figure 5.** Energy density as a function of C-rate associated with each Li$_x$V$_2$O$_5$ phase change for the o-IO and c-IO V$_2$O$_5$ Li battery cathodes grown by electrodeposition.

The drop in specific energy density is more pronounced when the rate is increased by an order of magnitude from 0.5 to 5 C, but further C-rate increases show limited changes to phase-related energy density. For the o-IO structure, the ω, γ, and δ phase energy densities decreased to ~73, 60 and 45 Wh kg$^{-1}$ respectively. However in the c-IO strucure, the ω, γ, and δ phase energy densities were suppressed to 31, 18, and 1.55 Wh kg$^{-1}$ respectively. In galvanostatic discharge the reduction in voltage at higher C-rate link with higher lithium mole fractions (Fig. 4 (b) and (d)), and so the c-IO network is less capable than the thinner o-IO at retaining a higher energy density at lower rates. The energy density of the c-IO requires discharge to the nominally the irreversible ω phase at all C-rates to maximize its energy density to values comparable to o-IO at low C-rates. As such, c-IO are less energy dense when discharged to higher cut-off potentials compared to o-IO cathodes.

The response and limits of the $V_2O_5$ IO structure to retain its capacity at successively higher C-rates was investigated by discharging and charging cells in two different experiments. First, we conducted the standard C-rate test using a single flooded cell with no prelithiation and no constant potential charging step in each cycle) over 10 cycles at C-rates ranging from 0.5 – 30 C, followed by a return to 0.5 C for 10 cycles. In the second experiment, we used freshly prepared o-IO of $V_2O_5$ in separate flooded cells, each separately discharged and recharged at each of the C-rates for 10 cycles each. Figure 6(a) shows the discharge curves for separate fresh o-IO $V_2O_5$ cathodes at each C-rates, and the discharge curves of a single o-IO cathode discharged at each C-rate successively.

Voltage step characteristics of $Li_xV_2O_5$ phases remain consistent up to 5 C but shift to lower potentials (also shown in in Figure 4) at higher rates when o-IO is discharged from pristine state at higher rates. Successive discharging to higher rates (Figure 6(a)) using the same electrode, clearly shows the marked suppression in specific capacity once the C-rate is increased to 5 C following cycling at 0.5 C. The capacity reduces to just under 50 mAh g$^{-1}$ (from over 310 mAh g$^{-1}$) and the voltage steps corresponding to phase changes during lithiation are absent from the discharge curve. Charge storage reactions are minimized to near-zero values as the rate in increased to 30 C for a single IO electrode after previous cycling at lower rates. Figure 6(b) shows the specific capacity retention over 10 cycles obtained on a single $V_2O_5$ o-IO structure for each C rate. The initial capacity decays at 0.5 C as expected from initial lithiation reactions from 310 mAh g$^{-1}$ to 110 mAh g$^{-1}$ after 10 cycles. Typically, $V_2O_5$ IO structured cathodes stabilize their capacity after 10-15 cycles as previously reported(8). Here, the capacity retention with cycling was stable for successive cycles at higher C-rate. For the following 40 cycles at 5 C to 30 C however, the o-IO capacity reduced from 30 mAh g$^{-1}$ to 8 mAh g$^{-1}$, and finally to 0.06 mAh g$^{-1}$, essentially inactive as an electrode at the fastest rate. The dramatic drop in capacity from 10 C to 30 C is consistent with the low lithium mole fraction obtained at these high rates, the corresponding lower voltage and energy density associated with the discharged $V_2O_5$ phase at 10 C or higher (*cf.* Figures 4 and 5).

Once the C-rate was reset to 0.5 C, the capacity recovered to a final 106 mAh g$^{-1}$ at the 50$^{th}$ overall cycle, almost a full recovery in specific capacity, from a near-zero specific capacity at 30 C. The SEM inset for Figure 5(b) shows the o-IO structure after the single cell was subjected to the successive C-rates over 50 cycles. Aside from slightly thicker IO walls from lithiation, the ordered porous structure remained intact.

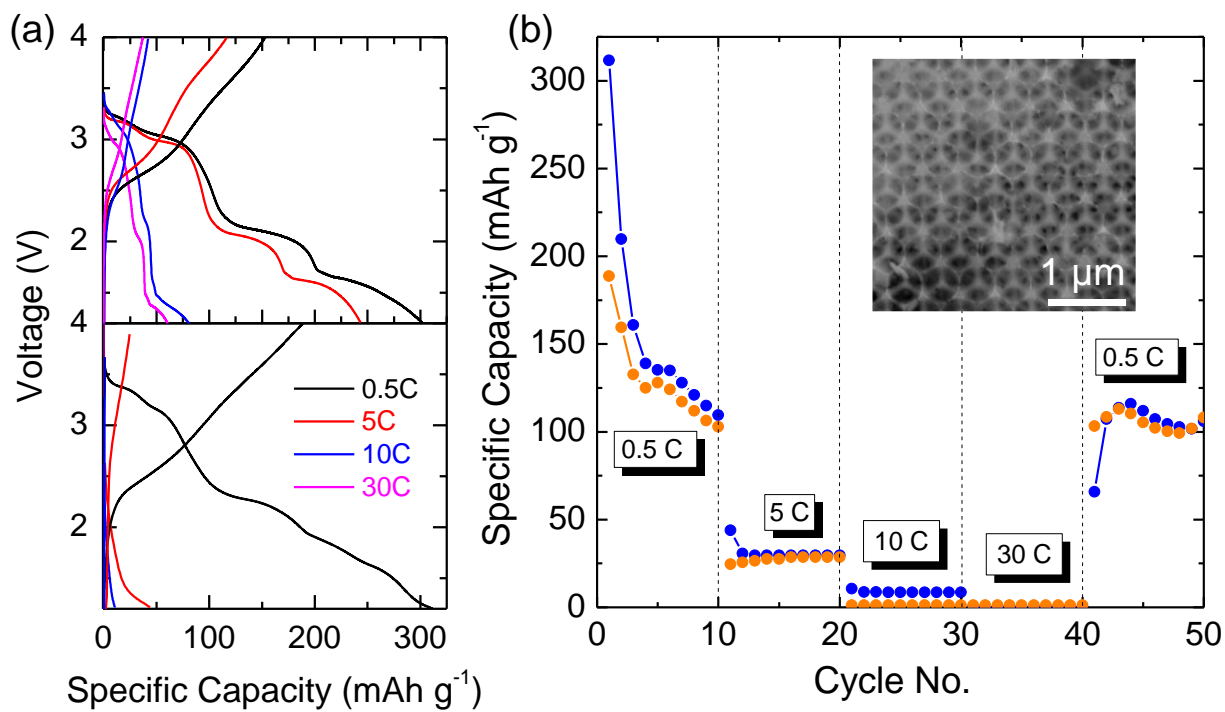

**Figure 6.** (a) First charge-discharge curves of separate o-IO $V_2O_5$ cathodes, each of which were discharged and charged at various C-rates. Underneath is the corresponding charge-discharge profiles of a single o-IO cathode successively discharged and charged at each C-rate. (b) A single o-IO $V_2O_5$ cathode cycled for 10 cycles at each C-rate for a total of 50 cycles, with SEM image of the material after cycling.

When $V_2O_5$ (in its orthorhombic α-$V_2O_5$ phase) is intercalated by Li$^+$, the resulting δ-$V_2O_5$ phase is known to cause an ~11% volume expansion, which is then reduced by over 6% as the structure contracts(78) to the γ-$V_2O_5$ phase, and these physical changes occur at rates where intercalation occurs (<10 C rate) and typically to discharge potentials >1.8 V at the slower rates. Clearly, a mechanism not related to lithiation dominates at higher rates even for interconnected and open porous material that are soaked with electrolyte in flooded cells, which we show later is related to the electrical conductivity from the current collector to the electrolyte. This mechanism is must be separate from electrochemical processes, as the phase and nature of the $V_2O_5$ is essentially unchanged due to the miniscule specific charge associated with negligible lithiation.

Next, we examined the rate response of both o-IO and c-IO $V_2O_5$ cathodes separately at each C rate over 25 cycles. Pristine electrodes were used in each case for each C-rate, to assess initial lithiation reactions and capacity decay, and the limits in rate response in both ordered macroporous materials. The discharge capacities shown in Figure 7(a) demonstrate several features. The initial specific capacity reduces with faster rate and values begin to level out beyond 10-15 cycles, once the reversible lithiated vanadate phase is formed. In most cases, the o-IO structures (open topped, thinner film) maintain a higher specific capacity during cycling, but we do note some variation in many cells we tested. More important is the effective suppression of the specific capacity during cycling at higher rates. The o-IO $V_2O_5$ structures had 25th cycle discharge specific capacities of 42, 25, 40, and 4 mAh g$^{-1}$, when cycled at the different C-rates of 0.5, 5, 10 and 30 C, respectively. Likewise, the c-IO $V_2O_5$ structures displayed similar reductions in final specific capacities of 10, 35, 30, and ~1 mAh g$^{-1}$. In all cases, the internal IO structure imaged by SEM after cycling (Figure 7(b)) is generally maintained after 25 cycles, with limited swelling and no obvious destruction of the macroporous order or interconnectivity.

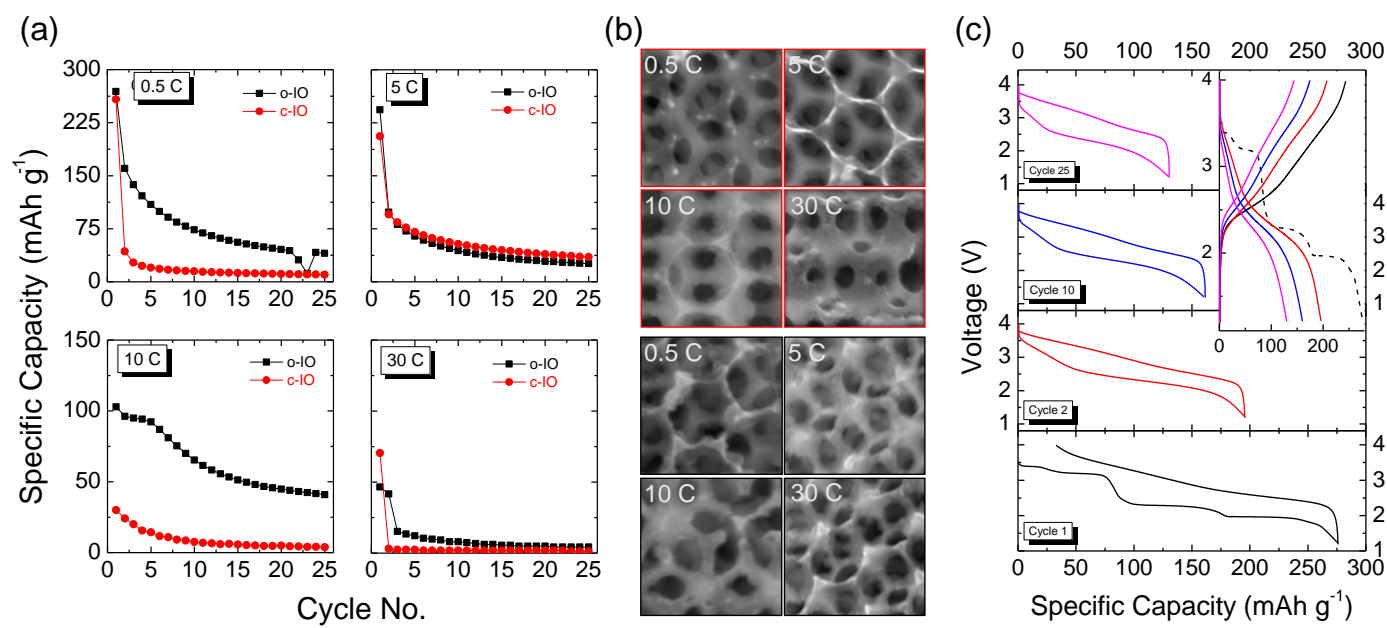

**Figure 7.** (a) Specific capacity from initial lithiation versus cycle number of o-IO and c-IO $V_2O_5$ electrodes in flooded 3-electrode Li battery cells at rates of 0.5, 5, 10 and 30 C. (b) Corresponding SEM images of the c-IO (red outline) and o-IO (black outline) morphology after 25 cycles. (c) Charge-discharge cycles (25 cycles) showing the voltage fade, cycle hysteresis, capacity fade for an o-IO electrode cycles at 0.5 C. No constant potential charging was applied to this cathode after the first galvanostatic charge.

In Figure 7(c) we show a series of charge-discharge cycles from the o-IO structure at the slowest C-rate (0.5 C). Apart from the specific capacity value differences, the response of the c-IO structure is nominally very similar. The initial discharge shows the phase transformation of $V_2O_5$ to its $\omega$-Li$_x$V$_2$O$_5$ phase. The largest capacity loss occurs in this first cycle, where after the first charge ~40 mAh g$^{-1}$ (14.5% reduction) is lost. Subsequent cycles of the lithiated $V_2O_5$ IO electrodes retain full (95-99%) Coulombic. The cycle hysteresis is apparent in the first cycle, but its form is nearly identical from the second cycle onwards. The voltage difference widens with cycling, indicating reduced energy/power density. There are several reasons for hysteresis in cathode materials, and based on microscopy and electrochemical data, we rule out dominant fracture or stress-related effects. Phase transformation effects are also eliminated as the primary transformations occurs during the voltage steps in the first discharge, and the voltage profiles are near-identical from cycle 2 onwards for both o-IO and c-Io electrodes at all C-rates (*cf.* $\frac{dQ}{dV}$ curves in Figure 4). Polarization concentration, particularly for semiconducting oxides such as $V_2O_5$ is not considered dominant as the lithiated phase has a lower, indirect bandgap, and the lithiation rate (specific current) is low. Internal resistance and lithium ion diffusion in the soaked electrolyte remain the primary candidates. Solid state diffusion or lithium gradient within the IO material itself is likely not an issue as the IO has thin macropore walls (~20 nm, within solid state diffusion length limits at 0.5 C rate) soaked with electrolyte in a flooded cell cycled at low rate. Thus, changes to out-of-plane electrical conductivity caused initial by the phase change to $Li_xV_2O_5$ are likely a significant contributor to the rapid onset of capacity fade to negligible values at rates > 10 C. The modified Butler-Volmer relation put forward by Lu et al.(79) similar to another based on chemical potential consideration by Bower et al.(80) relates the voltage gap at a fixed specific capacity (time) for an equal charging and discharging specific current. Bower's model elegantly relates the situation to plastic flow and diffusion. In our case, measurements confirm an overall diffusion-limited lithiation process. In the absence of stress contributions to reversible lithiation, their relation reduces to the form $\Delta V = \left(\frac{4RT}{F}\right)\sinh^{-1}\left(\frac{|i|}{2i_0}\right)$, where an overpotential is still defined with a magnitude of exactly $\frac{\Delta V}{2}$. The exchange current density takes the standard form assuming a transfer coefficient of 0.5, and that the lithium concentration at the surface of the cathode IO material and the concentration in the electrolyte do not vary significantly. We assume these conditions are satisfied and the electronic conductivity or out-of-plane electrode resistance (electron density at the IO surfaces to facilitate lithium reduction/intercalation) limits the high C-rate response of the IO structure, analyzed in more detail further on.

Before examining the basis for sever high rate capacity suppression, we also examined the phase and crystallinity changes to the IO cathode is the flooded Li battery cells. We have shown that IO structure is maintained, so evidence of lithiated vanadate phases should of course been measurable when normal specific capacity values are stored. When the capacity is suppressed due to near-zero reaction, we would expect to see little or no change to the original IO phase or crystallinity. This would confirm a mechanism related to cation/electron density and diffusion limitations. Raman scattering analysis in Figures 8(a,b) was acquired for both o-IO and c-IO electrodes after 25 cycles at each C-rate. Raman analysis of structural changes to $V_2O_5$ as a function of lithium mole fraction and C-rate has been examined in thin film and bulk electrodes,(81-84) leading to a good understanding of lithiated phases(85). After 25 cycles at 0.5 C, o-IO structure undergoes a blue-shift in the vanadyl (V=O) due to its reduction due to $Li_xV_2O_5$ and a suppression of the six-coordinated vanadium (V-O-V mode) is caused by the change to the lamellar structure of $V_2O_5$ after cation insertion and removal by cyclical lithiation(57).

The c-IO structure undergoes a significant structural change likely caused by bulk-level modifications from lithiation of the dense overlayer. In general, the c-IO structure retains some intensity of $V_2O_5$ modes when cycled at the higher rates, which is not too surprising as the film is bulk $V_2O_5$ that usually requires a longer time to fully lithiate compared to several-nm thin walls of the underlying electrolyte-filled IO. As the overlayer does not cover the IO completely, we expect that the c-IO is also filled with electrolyte in a flooded cell just as the o-IO is. However, as is well known from vanadium oxide thin film battery electrode investigations, the cation vdW gallery spacing is perpendicular to the ionic diffusion direction (i.e. layered structure is parallel to top surface of the o-IO and also the current collector), which normally impedes faster ion insertion. This observation is consistent with the lower lithium mole fraction we measured for c-IO as a function of C-rate. Consequently, a limited phase change in the c-IO structures is found compared to o-IO structure, in which the majority of the $V_2O_5$ modes have been irreversibly modified to a cycled lithiation vanadate after the first cycle. The V=O reduction to a charged V-O species at lower wavenumber, is also consistent with a higher uptake of lithium in the o-IO structures. This mode is erased once the lamellar and stoichiometric $V_2O_5$ phase has been irreversibly modified. High rate cycling of the c-IO structure retains this V=O feature, which is a fingerprint for very limited lithiation. Its presence is consistent with the electrochemical data shown earlier, confirming a limited lithium uptake and a supressed specific capacity at higher rates up to 30 C.

We also examined the c-IO and o-IO $V_2O_5$ macroporous structures by XRD to corroborate the changes in composition/lithium mole fraction to changes in crystalline phase. Figures 8(c, d) show the XRD patterns acquired from the electrodeposited o-IO and c-IO IOs after 25 cycles at the same C-rates. XRD patterns confirm crystalline orthorhombic $V_2O_5$ with a P*mmn* space group for both o-IO and c-IO(83). In both cases, the crystalline phase changes are consistent after cycling at each rate for 25 cycles. The orthorhombic $\alpha$-$V_2O_5$ structure is no longer detectable after 25 cycles at any C-rate. The lithiated vanadate retains the 110 and 200 reflections of the layered host $V_2O_5$, while all other reflections from the orthorhombic structure, namely those involving layer-to-layer coordination with the unit cells, have negligible intensity following cycling. We note that the intensity reduction is not due to delamination of material, nor consumption of cathode material during cycling form electrochemical processes involving electrolyte interphases, but from lithiation-induced phase change. This observation is interesting as it points to a sensitivity to reversible phase change even for comparatively low lithium concentration. Particularly at the higher C-rates, the c-IO structure achieves a maximum lithiated mole fraction of $\chi = 0.2$ in the initial discharge, and for the subsequent 25 cycles the capacity severely fades. At constant high specific current, the lithium concentration available to the electrode is greater than the lithium being intercalated since each pore within the IO is constantly filled with electrolyte in the flooded cell.

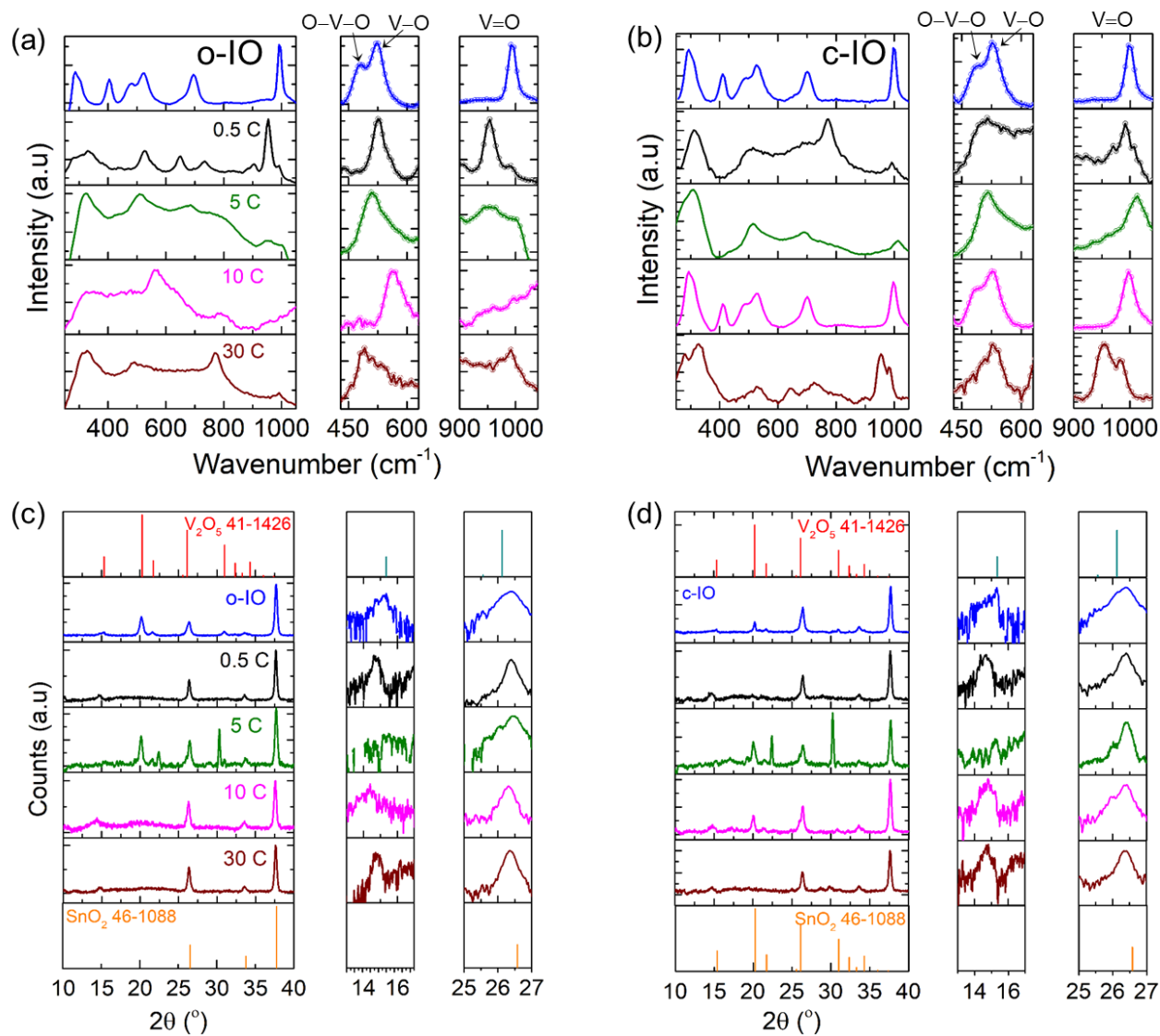

**Figure 8.** Raman scattering spectra of electrodeposited (a) o-IO $V_2O_5$ and (b) c-IO cathodes after 25 cycles at each C rate. Electrodes were analysed in ambient environment after removal from the electrolyte. The spectrum of the as-deposited $V_2O_5$ IO is also provided in each case *cf*. Figure 1 for spectral bond identification. (c) XRD patterns for as-deposited IO, o-IO and (d) c-IO $V_2O_5$ cathodes at each C rate. Reference patterns for $V_2O_5$ (PDF 41-1426) and $SnO_2$ (PDF 64-1088) are also shown. The $SnO_2$ reflections come from the FTO coating on the underlying substrates.

A permanent phase change to a lithiated vanadate is maintained during cycling. These analyses show some interesting features. Limited lithiation at faster C-rates is an effect we observe coincident with lower capacity. Unlike other systems using coin cell configurations and variation in slurry cast electrode thickness, we can ensure complete flooding of all pores, effectively minimizing Li electrolyte diffusion limitations. For any thickness of IO layer, the internal walls retain similar thickness as this is defined by the template sphere size. Hence, the effective thickness for lithiation is not enlarged using the c-IO, yet faster rates completely suppress the lithiation reaction. The capacity is fully recoverable when the C-rate is reduced and the internal porosity, materials (wall) thickness and structure remain similar. Nominally, thickness changes to an ordered porous IO would not worsen Li-ion diffusion rate *within* a pore, which should be similar everywhere once filled with electrolyte. The internal resistance and surface electron density will be affected by thickness via the ohmic drop from the current collector through the IO. A consistent observation is that fresh IO cathodes discharged separately at each C-rate, and a single cathode discharged at all rates, both show high rate capacity suppression. The non-lithiated electrodes are typically more electrically conductive that lithiated vanadate, which are known to become less conductive upon lithiation. In effect, fast C-rates for ordered macroporous electrodes in general, may depend on the intrinsic electrical conductivity even when their structure promotes fast rate capability (thin material, flooded electrolyte, porosity etc.).

*Chronoamperometric examination of IO electrode limits at fast C-rate*

We investigate next the effect of out-of-plane limited electrical conductivity using chronoamperometry (CA) to examine the supressed capacity in o-IO and c-IO at higher rates, while considering various conduction phenomena that occur within battery electrodes(86). Cell testing using galvanostatic mode at a range of specific currents for a fixed number of cycles is the standard approach but can take a long time. This can limit rapid assessment of new materials, electrode structures, or large electrode sets quickly. CA allows fast rate-dependent measurement of electrodes and has been shown to be fully consistent with a wide range of electrodes types, thickness slurry composition under galvanostatic testing(62, 63). The method involves a potential step to the lower cut-off voltage for the electrode, and a measurement of current transient with time at constant potential. Heubner et al. proposed a set of equations that allows conversion of this current transient directly to capacity – C-rate curves, giving high data point density across several orders of magnitude of C-rate in a matter of minutes.  Here, we use this method to compare to the standard galvanostatic discharge curves acquired at various C-rates.  The experimentally measured out-of-plane conductivity is typically in the $10^{-5} – 10^{-4}$ S cm$^{-1}$ range and does not vary by an order of magnitude between the ~6 μm thick o-IO and the ~12 μm thick c-IO.

There are several governing equations for the CA approach. One set developed by Tian and Coleman et al. alters that typical C-rate definition to an R-rate which is the specific current related to the experimental measured capacity (1/R being the discharge time) for a given electrode(14, 63). This is often more useful, but here we chose to use the CA method linked to C-rate reported by Heubner et al.(46), related maximum theoretical capacity since $V_2O_5$ behavior is well known and has been examined at specific current linked to C-rate more often in the literature. However, we should point out that an examination of the quality of fits using both approaches has shown that the best overall fitting to specific capacity vs rate curves is when the rate is defined as the R-rate. We will compare both approaches in this work while comparing to the galvanostatically obtained capacity vs rate data for o-IO and c-IOs. Using Refs. (63) and (14) as examples, readers can interchange between C-rate and R-rate and apply the governing equations to fit the capacity vs rate data accordingly to extract meaningful quantitative data from various electrode types using CA.

In brief, Heubner et al. proposed that the $I$(t) converts to C-rate according to(87)

$$\text{C-rate} = \frac{\frac{I}{M}}{\int_0^\infty \left(\frac{I}{M}\right) dt} \quad (1)$$

where $\frac{I}{M}$ is measured current normalised by electrode mass. The specific current is then normalised to the specific current after infinite time, implying that the final specific charge approaches the theoretical maximum for the electrode. This assumes the final specific capacity measured from the transients and the theoretical specific capacity are equivalent. Experimentally, we determine this value by limiting the potential step measurement over a time where the current becomes very small. This value is the experimental maximum capacity of the electrode using CA and comes close to the theoretical maximum capacity at very low rates. The corresponding definition of R-rate (normalized to the actual measured maximum low rate capacity over a fixed time *t*, is

$$\text{R-rate} = \frac{\frac{I}{M}}{\int_0^t \left(\frac{I}{M}\right) dt} \quad (2)$$

These equations transform the current vs time data obtained a potential step to the lower cut-off potential, to the specific capacity vs C-rate (or R-rate). Then, a comparison can be made to galvanostatic rate-dependent data shown earlier for $V_2O_5$ o-IO and c-IO electrodes. Detailed in Refs (14) and (63), the CA data can be fit to extract parameters such as charging time and the effect of electrical conductivity on the charging rate, among other useful parameters. Both models are used here to compare to the galvanostatic rate data acquired for our flooded IO electrodes in Li battery cells, with fits described by

$$\frac{Q}{M} = Q_M[1 - 2(\tau R_C)^n] \quad (3)$$

for C-rate ($R_C$) dependency and

$$\frac{Q}{M} = \frac{Q_M}{1 + 2(R\tau)^n} \quad (4)$$

Here, $\tau$ represents a general characteristic time associated with charging and discharging, and $Q_M$ is the low rate specific capacity. As detailed elsewhere(14), an exponent $n$ is introduced to generalise the fit equations 3 and 4 to allow for charge storage processes that range from diffusion controlled to kinetically controlled, i.e. $0.5 \leq n \leq 1$.

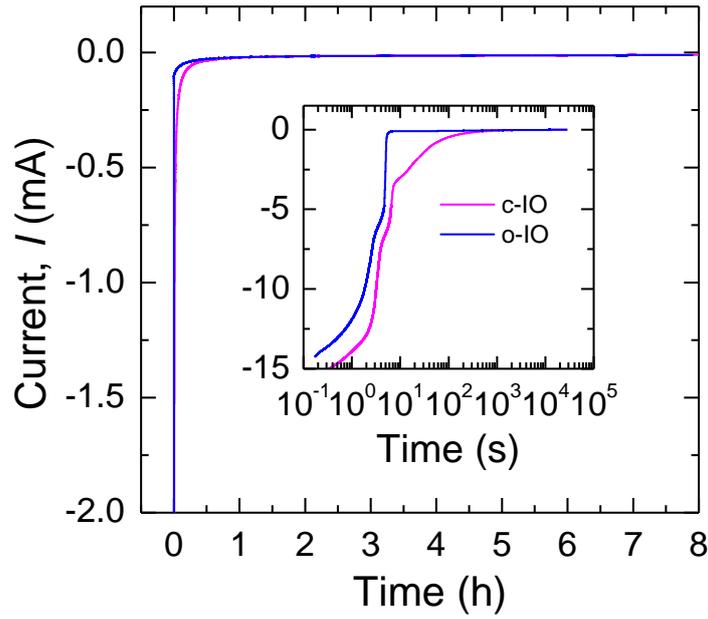

**Figure 9.** Potentiostatic $I$(t) transients acquired for o-IO and c-IO $V_2O_5$ Li battery electrodes in a flooded cell after potential step to the lower cut-off voltage of 1.2 V (Li$^+$/Li) acquired over a 6-hour period. Both o-IO and c-IO electrodes were used as-made, with no prelithiation, discharge or cycling history.

Figure 9 shows the $I$(t) potential step curves for the o-IO and c-IO electrodes in the flooded Li battery cell, where the potential was stepped from open circuit to the lower cut-off potential and held constant. The CA $I$(t) transients were taken from fresh electrodes, and it is interesting to see the fast and slow changes to the current (plateaus) in the transient are also found here as are commonly observed for $V_2O_5$ in galvanostatic data during lithiation of $V_2O_5$. These curves were converted to $\frac{Q}{M}$ vs C-rate and $\frac{Q}{M}$ vs R-rate curves in Fig. 10(a,b). Clearly, using the model of Huebner et al. shows a much faster reduction in relative specific capacity with C-rate compared to Tian's model. When these curves are compared to the measured average specific capacity values as a function of rate in Figure 11(a,b) for c-IO and c-IO electrodes, significant differences in values are found. The CA method allows rate analysis for values up to ~$10^6$ C or more, and so the majority of practicable rates (and those studied here) are found in the early regions of this curve. This data is acquired very quickly during the current transient following the potential step. In Fig. 10, we show the average specific capacity as a function of C-rate for two cases: (1) where a fresh o-IO and c-IO electrode is used for each C-rate, and (2) where o-IO and c-IO electrodes are used to acquire capacity cumulative cycling data at all rates, as shown in Fig. 6 for example.

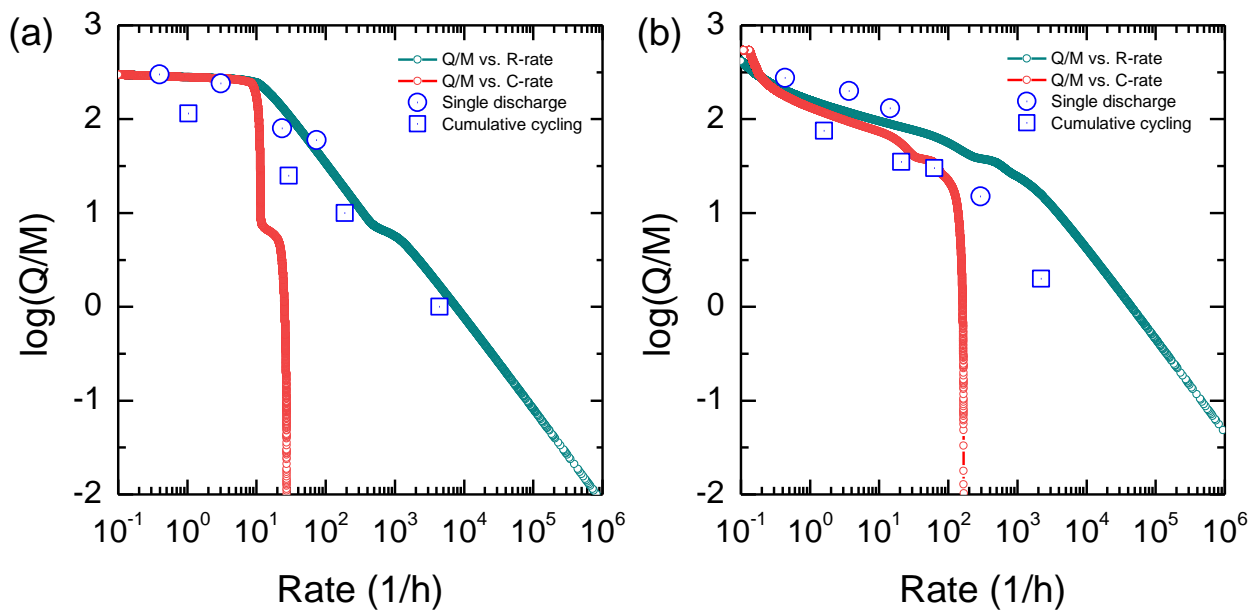

**Figure 10.** Q/M plots for o-IO and c-IO $V_2O_5$ Li battery electrodes in a flooded cell after potential step to the lower cut-off voltage of 1.2 V (Li$^+$/Li) as a function of C-rate and R-rate. Specific capacity values extracted from galvanostatic discharge curves of o-IO and c-IO electrodes (e.g. refer to Figs 3 and 6) are independently (not fitted) overlaid on the Q/M plots. Two different measurements of specific capacity data for o-IO and c-IO $V_2O_5$ electrodes are shown: (1) fresh unlithiated electrodes each discharged at 0.5 C, 5 C, 10 C, and 30 C (1$^{st}$ discharge). (2) A single o-IO or c-IO electrodes that underwent 25 cycles at each of the four C-rates. The plots were also fitted using Eqs 3 and 4.

The data in Fig. 10 confirm that the CA method predicts rate-behavior with respect to theoretical capacity (C-rate), or maximum measured capacity at the lowest rate (R-rate) with high resolution. Direct comparison with charge-discharge curve measurements show very good agreement for o-IO and c-IO electrode single discharge data. We should point out that the CA potential step was also acquired from fresh electrodes. Even though $V_2O_5$ undergoes well-defined steps in potential under constant current lithiation, we observe definite plateaus in the CA curves. This phenomenon was just recently observed for slurry cast graphite and NCA electrodes of various thickness(62). The plateau was linked to a change from diffusion-limited lithiation, to a high rate region that was limited by electrical conductivity of the electrode. For our electrodes, the GCD data tracks the R-rate CA curve better than the C-rate curve and the highest rate (30 C) data where the capacity is suppressed, occurs after the knew in the curve, indicating electrical conductivity limitations. Unlike previous CA analyses, our electrodes are formally interconnected, filled with electrolyte (flooded) and devoid of other materials.

In previous work, we examined cyclic voltammetric response of $V_2O_5$ IO electrodes(9) and found that above 50 mV s$^{-1}$ scan rate (which corresponds to the kink feature in the (red) C-rate curve in Figs 10(a) and (b)), the cathodic peaks associated phase change lithiation of the $V_2O_5$ (voltage steps in the galvanostatic curve) disappear during a voltammogram. This is also found in nanomaterials of $V_2O_5$ (88) at higher scan rates. At high rates, the $V_2O_5$ IO voltammogram curve looks capacitive in nature. In Fig. 10(a), the o-IO electrode shows a fall off at ~10 C, consistent with data of separate cells in Fig. 3 for example. For the c-IO, the fall of is more gradual as the rate is increased, also consistent with the CA data in Fig. 10 (b). Using the dependence from Tian et al. in Eq 4, the time associated with discharge, $\tau$ can be related to rate by $R_T = (1/2)^{1/n}/\tau$, and as this correlates with the high rate kink in the CA curve, we demonstrate that it is observable for materials with much lower conductivity (over 5 orders of magnitude) that previously thought. From the potentiostatic transients, the transition from high rate to low rate behavior occurs from 90-100 s after potential step, after which the decreasing current describes the low rate behaviour prior to the kink features in the curves in Fig. 10. Using a value of 90 s, the rate is predicted as ~10 C (where *n* = 0.5), very close to the experimental observations by normal discharge measurements.

If we compare to the change in lithiated mole fraction in Fig. 4 and the electrode energy density shown in Fig. 5, the overall trend is similar; a large reduction occurs at rates > 10 C. For the c-IO, this rate is slightly larger at ~25 C. Above these rates, the electrodes stores essential no useful capacity and so limited lithiated mole fraction, negligible change to crystal structure/phase, suppressed energy density due to a lower potential associated with reaction for various vanadate phases. Dialling back the rate shows a full recovery in specific capacity through intercalation other that background capacitive charge common to all material polarised in an electrolyte.

A separate analysis that focused on the well-known diffusion limitation of the cation in the electrolyte by Heubner et al.(61) assigns the sever capacity decay to a C-rate that exceeds a diffusion-limited current that is related to C-rate by the areal capacity, also explored by Park et al. for standard slurry electrodes where the areal capacity and electrode thickness are sensitive to rate (17). In Ref. (61) for example, tortuosity are microporosity more common to slurried electrodes with various thickness, and electrolyte depletion, influence rate behaviour. Macroporous IOs (porosity on the hundreds of nm scale that is approximately constant for all thickness) with flooded electrolyte is more affected by material conductivity at higher rates. The effect noticed by Heubner is similar at high specific current above their DLC – capacity is essentially suppressed. However, the link to electrical conductivity was not formally defined and the dependence on binder content, particle size and conductive additive content are less relevant to our electrode designs. Of course, the increased porosity (reducing electrode density) is usually described to be beneficial for higher rate applications. Cleary, this porosity effect is not ubiquitous and is based on slurried electrode with conductive additive and larger pores. If the intrinsic electrical conductivity out-of-plane requires a high-volume fraction of graphite additive, gravimetric capacity optimization is not straightforward. IOs are interconnected and continuously conductive along the resistive path to the surface (all surfaces in the 3D IO) where Li intercalation occurs. The rate will depend on electron density at that surface once all other influencing parameters are not limiting (e.g. electrolyte Li-ion diffusivity or concentration, among others).

**Conclusions**

We have shown that interconnected ordered macroporous structures of $V_2O_5$ as a working example, have intrinsic limits to performing well at high charge and discharge rates when used in lithium batteries. When these electrodes are tested in Li battery flooded cells, fully soaked with electrolyte and without any additives such as binder or graphitic materials, they show complete specific capacity suppression at rates greater than 10 C. Unlike slurry-cast composite electrode with more complex microporosity, inverse opals electrodes grown by electrodeposition (and some other methods) typical perform quite well for a single phase porous material at low rates and maintain electrical interconnection and the same porosity irrespective of thickness. At higher rates, the electrical conductivity (perpendicular to the current collector) limits the rate at which reasonable charge is stored by intercalation.

For thicker macroporous electrodes, the porosity is nearly identical since it is defined by the opal photonic crystal template. The two electrodes studied here (one ~ 6 µm thick, the other ~12 µm thick), both are filled with electrolyte such that depletion from bulk electrolyte is not an issue as it is for thick calendered electrodes. However, as the intrinsic measured conductivity is similar for both materials when measured in the dry ambient prior to lithiation, ~4-6 order of magnitude lower than many slurried electrodes with ~5-10% graphitic additives, both thin and thick electrode show complete capacity suppression above 10 C. This capacity can be completely recovered when the rate is reduced to 0.5 C. Examination of the phase changes of $V_2O_5$ are fully consistent with expected lithiation mechanisms. In the low rate region, we showed using XRD, SEM and Raman scattering, that phase changes and limited structural changes occur after cycling and can correlate the specific capacity to $Li_xV_2O_5$ phases. In the higher rate region where negligible charge is sored, the material remains relatively pristine indicating that no reaction with Li occurs. Thus, when the rate is reduced to 0.5 C in this case, the only cycling history where lithiation occurs are for cycling period at rates < 10 C.

We also used recently developed methods to quantify rate dependence in composite slurry Li-ion electrodes. This approach uses a potential step to the lower cut-off potential of the electrode and provides capacity vs rate data much more quickly and with much higher data resolution than standard galvanostatic cycling. By comparing specific capacity acquired from standard galvanosatic discharge-charge cycling to the chronoamperometric analysis, the data confirm an electrical limitation to high rate response where no charge is stored, and predicts the C-rate below which the IO electrode undergoes intercalation reactions. This work may be generally applicable to many more macroporous ordered electrodes, and suggest that higher conductivity materials are necessary to ensure faster rate behaviour in battery cells. One general comment is that porosity (with minimal cation tortuosity) is important for rate behavior modification, along with the other benefits for some material that undergo a lot of expansion during cycling and for limiting solid state diffusion issues (akin to nanoscaling materials). Even if cation diffusivity limitations are removed, electrical conductivity remains important for higher rates. Slurry composites are more sensitive to rate at higher thickness, particularly for areal capacity optimization, and microporosity, additives, and diffusion of cation within electrolyte and various sizes of random particulates play a role. IOs are 'cleaner' from this perspective, but are much less energy dense due to the porosity. However, as a model system, they provide useful information on the limitations to high rates when tested in a flooded Li battery cell.

**Acknowledgments**

This work was supported by a research grant from Science Foundation Ireland (SFI) under Grant Number 14/IA/2581. This publication has emanated from research conducted with the financial support of Science Foundation Ireland (SFI) and is co-funded under the European Regional Development Fund under the AMBER award, Grant Number 12/RC/2278_2.